\documentclass{emulateapj}
\usepackage{apjfonts}
\usepackage{bm}
\usepackage{amsmath}
\usepackage{rotating}

\newcommand\lsim{\mathrel{\rlap{\lower4pt\hbox{\hskip1pt$\sim$}}
        \raise1pt\hbox{$<$}}}
\newcommand\gsim{\mathrel{\rlap{\lower4pt\hbox{\hskip1pt$\sim$}}
        \raise1pt\hbox{$>$}}}
\newcommand{\D}{\mathrm{d}}

\newcommand{\yr}{\mathrm{yr}}

\newcommand{\Hz}{\mathrm{Hz}}

\newcommand{\mHz}{\mathrm{mHz}}
\newcommand{\Mpc}{\mathrm{Mpc}}

\newcommand{\dL}{d_{\rm L}}
\newcommand{\isco}{\rm isco}

\shorttitle{Finding the Electromagnetic Counterparts of Standard Sirens}
\shortauthors{Kocsis, Frei, Haiman, \& Menou}

\begin{document}

\title{Finding the Electromagnetic Counterparts of Cosmological
Standard Sirens}

\author{Bence Kocsis}
\affil{Institute of Physics, E\"otv\"os University, P\'azm\'any P.
s. 1/A, 1117 Budapest, Hungary; bkocsis@complex.elte.hu}
\author{Zsolt Frei}
\affil{Institute of Physics, E\"otv\"os University, P\'azm\'any P.
s. 1/A, 1117 Budapest, Hungary; frei@alcyone.elte.hu}
\author{Zolt\'an Haiman}
\affil{Department of Astronomy, Columbia University, 550 West
120th Street, New York, NY 10027; zoltan@astro.columbia.edu}
\author{Kristen Menou}
\affil{Department of Astronomy, Columbia University, 550 West
120th Street, New York, NY 10027; kristen@astro.columbia.edu}

\submitted{Accepted for publication in \apj}

\begin{abstract}
The gravitational waves (GW) emitted during the coalescence of
supermassive black holes (SMBHs) in the mass range $\sim
(10^4$--$10^7)\,{\rm M_\odot}/(1+z)$ will be detectable out to high
redshifts with the future {\it Laser Interferometric Space Antenna}
({\it LISA}). The distance and direction to these ``standard
sirens'' can be inferred directly from the GW signal, with a
precision that depends on the masses, spins and geometry of the
merging system. In a given cosmology, the {\it LISA}-measured
luminosity distance translates into a redshift shell.  We calculate
the size and shape of the corresponding three--dimensional error
volume in which an electromagnetic counterpart to a {\it LISA} event
could be found, taking into account errors in the background
cosmology (as expected by the time {\it LISA} flies), weak
gravitational lensing (de-)magnification due to inhomogeneities
along the line of sight, and potential source peculiar velocities.
Weak lensing errors largely exceed other sources of uncertainties
(by a factor of $\sim 7$ for typical sources at $z=1$). Under the
plausible assumption that SMBH-SMBH mergers are accompanied by gas
accretion leading to Eddington-limited quasar activity, we then
compute the number of quasars that would be found in a typical
three--dimensional {\it LISA} error volume, as a function of BH mass
and event redshift. Low redshifts offer the best opportunities to
identify quasar counterparts to cosmological standard sirens. For
mergers of $\sim(4\times10^5 - 10^7) {\rm M_\odot}$ SMBHs, the {\it
LISA} error volume will typically contain a single near-Eddington
quasar at $z\sim 1$. If SMBHs are spinning rapidly, the error volume
is smaller and may contain a unique quasar out to redshift $z\sim
3$. This will allow a straightforward test of the hypothesis that GW
events are accompanied by bright quasar activity and, if the
hypothesis proves correct, will guarantee the identification of a
unique quasar counterpart to a {\it LISA} event, with a B-band
luminosity of $L_B\sim (10^{10}-10^{11}) {\rm L_\odot}$. Robust
counterpart identifications would allow unprecedented tests of the
physics of SMBH accretion, such as precision--measurements of the
Eddington ratio. They would clarify the role of gas as a catalyst in
SMBH coalescences, and would also offer an alternative method to
constrain cosmological parameters.
\end{abstract}


\section{Introduction}
\label{sec:Introduction}

One of the main objectives of the {\it Laser Interferometric Space
Antenna} ({\it LISA}), to be launched around the year 2013
\citep{danz04}, is to detect the gravitational wave (GW) signals
associated with coalescing supermassive black holes (SMBH) at
cosmological distances. The {\it LISA} detector is designed to be
particularly sensitive in the frequency range between $(3\times
10^{-5}-10^{-4})\Hz\lsim f \lsim 0.1 \Hz$, allowing the detection of
binary coalescences with total masses between $10^4$ and $10^7\,{\rm
M_\odot}$ out to high redshifts.  The limiting redshift depends on
several factors (such as the orientation of the spins and orbital
plane of the SMBH binary, and its location on the sky relative to {\it
LISA}), and is expected to lie between $z\sim 5-10$ \citep{hou02}.  A
comparison of the gravitational waveform with the anticipated detector
noise can be used to estimate the accuracy with which {\it LISA} will
be able to extract the physical parameters of the coalescence events
\citep{hou02,bc04,vec04,hh05}. Of particular interest, in the context
of searching for electromagnetic (EM) counterparts, is whether the
spatial location of the GW event can be localized to within a
sufficiently small three-dimensional volume.  In this paper, we
determine the probability of finding a unique EM counterpart within
the expected error volume associated with SMBH merger events, for a
range of masses and redshifts.

It has been argued by \cite{vec04} that the identification of such EM
counterparts will be difficult because, in typical cases, there will
be too many counterpart candidates to choose from.  However,
\cite{vec04} associated counterparts with host galaxies and galaxy
clusters, and used only the 2D angular positioning information for the
analysis. In contrast, here we account for the 3D spatial information
by using the redshift of an electromagnetic counterpart candidate in
relation with the luminosity distance determined by {\it LISA} and we
focus on quasars as plausible counterparts. With these specifications,
we shall demonstrate that in some cases, a specific counterpart
can be uniquely determined.

If electromagnetic counterparts to {\it LISA} events exist, they will
likely be related to the accretion of gas onto the SMBHs involved in
the coalescence. Provided this accretion is not supply-limited, bright
quasar counterparts approaching the Eddington luminosity would then be
expected. A few additional arguments favor this scenario: galaxy
mergers in hierarchical scenarios of structure formation are expected
to deliver a significant amount of gas to the central regions of the
merging galaxies \citep{bh92}, and this gas may play a catalyst role
in driving SMBH coalescence \citep{bbr80,gr00,elcm04a}. Ultimately,
however, many of the complex processes involved remain poorly
understood. For example, \cite{an02} have argued that, in the limit of
a small mass ratio of the two SMBHs, a prompt and luminous
electromagnetic signal may be expected during coalescence, while
\cite{mp05} have argued that in the limit of equal mass SMBHs, only a
much delayed electromagnetic afterglow would be expected. All
``cosmological standard sirens'' may thus not be equal in their
potential for electromagnetic counterparts.
\footnote{The name ''standard sirens'' was suggested by Sterl Phinney
and Sean Carroll \citep{hh05}.}
Our working assumption in
the present study is that bright quasar activity is a plausible
electromagnetic counterpart to {\it LISA} events. This allows us to
quantify the feasibility of an unambiguous identification of such a
counterpart.  As we shall see below, the search for the counterparts
will allow a test of the assumption, as well.

The secure identification of the EM counterpart to even a single GW
event could be useful in different ways: (1) to improve our
understanding of the SMBH accretion physics, (2) to clarify the role
of gas as a catalyst in SMBH mergers and (3) to supply an independent
constraint on the background cosmology.  A joint GW -- EM analysis
could, in principle, determine the masses and orbital parameters of
the SMBH binary, and yield a precise measurement of the Eddington
ratio, $L/L_{\rm Edd}$, which will supply a key parameter in studies
of the evolution of the BH/quasar population
\citep{sb92,hl98,kh00,hm00,wl03}. This parameter is currently poorly
known (constrained by indirect empirical correlations;
\citealt{ves04}, \citealt{wu02}, \citealt{kas00}). The values range
from $\approx 0.1$ to $\gsim 1$, with indications that higher-z
quasars may be closer to $L_{\rm Edd}$ than the $z\sim 0$ quasars.
Likewise, a joint GW -- EM analysis could, in principle, be used to
estimate cosmological parameters \citep{s86}, by comparing the
luminosity distance (which is a direct observable by GWs) with the
redshift (as inferred from the spectrum of the counterpart -- in this
case, a quasar).  This would serve as a complement to constraints from
the luminosity distance to high-$z$ type Ia Supernovae (SNe), but with
different systematic errors, and with the potential of extending to
higher redshifts. New constraints, spanning the range $0<z<2$, would
be particularly well--suited to probe the properties of dark energy,
which is expected to become dynamically dominant within this cosmic
epoch.

The rest of this paper is organized as follows.
In \S~\ref{sec:LISAerrors}, we summarize our method to estimate the
angular and radial positioning errors expected from {\it LISA}, for
SMBHs with a range of masses at different redshifts.
In \S~\ref{sec:DLerrors}, we discuss the conversion of the luminosity
distance, as determined by {\it LISA} from the GW signal alone, to the
redshift of the source. In particular, we discuss the uncertainty in
the resulting redshift estimate.
In \S~\ref{sec:QSOLF}, we discuss our estimates for the number of
quasars that may be found in the 3D error volume provided by {\it
LISA}, based on the luminosity function and clustering properties of
known optical quasars.
In \S~\ref{sec:results}, we present our main results, and show that
for typical low-redshift GW events discovered by {\it LISA}, a unique quasar
counterpart may be identified.
In \S~\ref{sec:discussion}, we point out various implications of a
successful identification and discuss several caveats to this
conclusion.
Finally, in \S~\ref{sec:conclusions} we summarize our conclusions.
Unless stated otherwise, throughout this paper we assume a standard
cold--dark matter cosmology ($\Lambda$CDM), with ($\Omega_\Lambda$,
$\Omega_M$, $\Omega_b$, $H_0$) = (0.70, 0.30, 0.047, 70 km s$^{-1}$
Mpc$^{-1}$), consistent with the recent results from the {\it
Wilkinson Microwave Anisotropy Probe (WMAP)}\citep{spergel03} and the
Sloan Digital Sky Survey (SDSS) \citep{tegmark04}.

\section{Localizing {\it LISA} events}
\label{sec:LISAerrors}

A few studies have been carried out so far to address how accurately
{\it LISA} will measure the source parameters of a coalescing pair
of SMBHs.  In general, the accuracy depends on a large number of
parameters: a total of 17 parameters in the most general case
include 2 red-shifted mass parameters, 6 parameters related to the
BH spin vectors, the orbital eccentricity, the luminosity distance,
2 angles identifying the position on the sky, 3 angles that describe
the orientation of the orbit, a reference time, and a reference
phase. Due to the resulting computational limitations, various
studies have concentrated on small portions of the parameter space.
The most up--to--date calculations estimating parameter
uncertainties for SMBH in-spirals have been carried out by
\cite{bbw05a}; \cite{hh05}; and by \cite{vec04}. As compared to
previous studies, \cite{vec04} accounts for the effects of spins,
and shows that parameter estimation errors improve significantly (by
a factor of 3--10 for high spins) for selected parameters.
\cite{vec04} also adopts an optimistic {\it LISA} sensitivity curve,
by adopting the smallest observable frequency to be $\sim 3$ times
lower than previous estimates and only considers cases with equal
mass SMBHs. Our analysis, which relies on Vecchio's estimates, is
therefore approximate to this extent.

\begin{deluxetable}{lrrrr} \tablecolumns{6}
\tablewidth{0pt} \tablecaption{\label{tab:LISA}{\it LISA}
Measurement Errors} \tablehead{\colhead{} & \colhead{$\delta\cal
M/\cal M$} & \colhead{$\delta \mu/\mu$} & \colhead{$\delta \dL/\dL$}
& \colhead{$\delta\Omega$}} \startdata
   best     & $0.8\times10^{-5}$    &  $2\times10^{-5}$  &   $2\times10^{-3}$    &    $0.01 \deg^2$ \\
   typical  & $2\times10^{-5}$      &  $9\times10^{-5}$  &   $4\times10^{-3}$    &    $0.3 \deg^2$  \\
   worst    & $0.8\times10^{-3}$    &   0.1              &   $2\times10^{-2}$    &    $3 \deg^2$
\enddata
\tablecomments{Assumed SMBH binary parameters: $m_1=m_2=10^6 {\rm
M_\odot}$ and $z=1$.}
\end{deluxetable}

For concreteness, we adopt the parameter uncertainties obtained by
\cite{vec04} for an equal-mass SMBH binary with $m_1=m_2=10^6{\rm
M_\odot}$ at redshift $z=1$. The uncertainties vary as a function of
the fiducial orientation of the source relative to {\it LISA}, and
are primarily influenced by the BH spin magnitudes (i.e. higher
spins lead to smaller uncertainties). Here we distinguish three
cases. For our ''best'' case, we adopt the errors that correspond to
the 10th percentile of the distribution of uncertainties obtained by
\cite{vec04} for high BH spins (with dimensionless spin
$a=S/M^2=0.9$, where $S$ is the magnitude of the total spin and $M$
is the total mass). For our ''typical'' case, we adopt the errors
corresponding to the 50th percentile in the case of moderate BH
spins, with $a=0.5$. For our ''worst'' case, we adopt the 90th
percentile of the no spin case ($a=0$). Note that SMBHs are
generally expected to be spinning fast (e.g. \citealt{vol05}), so
that our ``best'' case may actually be representative of a fair
fraction of events. In Table~\ref{tab:LISA}, we list the errors on
the chirp and reduced masses, ${\cal M}=(m_1
m_2)^{3/5}/(m_1+m_2)^{1/5}$ and $\mu=(m_1
m_2)/(m_1+m_2)$
respectively, and on the GW source location (i.e. luminosity
distance, $\dL$, and solid angle, $\Omega$), for these three cases
of interest.

\begin{figure}
\centering\mbox{\includegraphics[width=8.5cm]{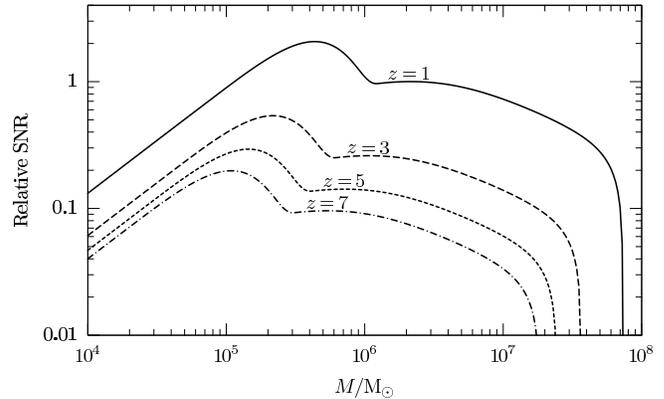}}
\caption[]{\label{fig:snr} Relative signal--to--noise ratio (SNR)
for {\it LISA} detections of the in-spiral phase of equal--mass
supermassive black hole coalescences, as a function of total mass,
$M$, and redshift, $z$.  A 1--year observation is assumed and the
normalization is for $M= 2\times 10^6~{\rm M_\odot}$ and $z=1$.}
\end{figure}

In \cite{vec04}, parameter errors have only been estimated for a
single choice of an equal--mass SMBH binary with total mass
$M_0=m_1+m_2=2\times 10^6 {\rm M_\odot}$ and redshift $z_0 = 1$.
Starting from this result, we crudely estimate the uncertainties
$\delta\dL$ and $\delta \Omega$ for other combinations of masses and
redshifts as follows.  First, note that the luminosity distance is
simply proportional to the inverse of the signal amplitude. Therefore
its estimator depends primarily on the total signal power, rather than
on the specific shape of the signal waveform\footnote{The signal power
also scales with the red-shifted chirp mass as ${\cal
M}_z^{5/6}$. However, this parameter can be determined independently
to high precision, from the phase information.}. The luminosity
distance error would then obey the simple scaling
\begin{equation}\label{eq:ddL(M,z)}
\frac{\delta\dL(M,z)}{\dL(M,z)} = \left[\frac{SNR(M,z)}{SNR(M_0,z_0)}\right]^{-1}
\frac{\delta\dL(M_0,z_0)}{\dL(M_0,z_0)}
\end{equation}
where $SNR(M,z)$ is the expected value of the signal to noise ratio
of the detection,
\begin{equation}\label{eq:SNR}
SNR\,^2(M,z) = 4\int_{f_{\rm a}(M_z,\Delta T)}^{f_{\isco}(M_z)}
\frac{h^*(f,M_z,z)h(f,M_z,z)}{S_n(f)}\D f.
\end{equation}
Here, $M_z=(1+z)M$ is the red-shifted total mass, $h(f,M_z,z)$ denotes
the Fourier decomposition of the signal detected by {\it LISA}, and
$S_n(f)$ is the RMS noise density per frequency interval (including
instrumental and confusion noise, \citealt{bc04}\footnote{For the
instrumental noise, instead of the approximation of \cite{bc04}, we
use the more exact sensitivity curve available at
http://www.srl.caltech.edu/\~~shane/sensitivity/.}). A crucial
parameter for high $M_z$ values inscribed in $S_n(f)$ is the low
frequency wall of the detector, which is further discussed below.  The
integration bound $f_{\isco}(M_z)$ corresponds to the innermost stable
circular orbit (ISCO), beyond which the gravitational waveform is not
well known, and $f_a(M_z,\Delta T)$ is the arrival frequency a time
$\Delta T$ before the ISCO is reached by the coalescing binary (see
\citealt{vec04} for details). Throughout this paper, we fix the
observation time of SMBH binaries at $\Delta T = 1$~yr, unless a
binary is so massive or at such a high redshift that it is not
observable by {\it LISA} for a full year and then $\Delta T < 1$~yr.
Note that $h(f,M_z,z)$ depends on other parameters, such as the
angular momentum vector orientation relative to {\it LISA}'s arms, the
magnitude of spins, etc. (see \citealt{vec04} and references therein).
For this estimate we calculate the $SNR$ with the leading order
(i.e. Newtonian) contribution.  The resulting dependence of the
signal--to--noise ratio on BH masses and redshift are shown in
Figure~\ref{fig:snr}.  The figure shows that the sensitivity degrades
significantly for distant sources, and also that it peaks in the mass
range of $10^5-10^6~{\rm M_\odot}$, which produce GWs near the optimal
frequencies within the {\it LISA} band.

The other important parameter, the angular position, is extracted
from the change in the relative orientation of {\it LISA} during its
orbit around the Sun. Since {\it LISA}'s orbital time--scale is much
longer than the inverse of the signal frequency, it is plausible to
assume that the angular position uncertainty decouples from the
intricate waveform, and improves linearly with the signal amplitude.
This assumption is consistent with Fig. 1 of \cite{bbw05b}
which shows that the inclusion of spin-orbit and spin-spin terms
modifies the waveforms but does not alter the angular resolution.
In fact, the angular resolution is shown to be independent even for
alternate theories of gravity (e.g. scalar-tensor and massive
graviton theories). Thus, similar to equation~(\ref{eq:ddL(M,z)}), we
estimate the mass and redshift dependence of the positioning
solid-angle error as
\begin{equation}\label{eq:dOmega(M,z)}
\delta\Omega(M,z) =
\left[\frac{SNR(M,z)}{SNR(M_0,z_0)}\right]^{-2}
\delta\Omega(M_0,z_0),
\end{equation}
where the $-2$ exponent assumes that the uncertainty in this
two--dimensional quantity is the product of independent
uncertainties in the one-dimensional azimuthal and polar angles.

A limitation of the above analysis at the high--mass end of the range
of SMBHs is that these events may not be visible for a full year, due
to the low--frequency noise wall, below which {\it LISA} looses
sensitivity. For instance, for $M_z>9\times 10^6 {\rm M_{\odot}}$
($1.3\times10^6 {\rm M_{\odot}}$), the low frequency noise wall at
0.03mHz (0.1mHz) is crossed less than $0.25\yr$ before reaching the
ISCO. At higher masses or redshifts, therefore, the angular
information, which is inscribed in a modulation with a $1\yr$ period,
becomes significantly harder to disentangle from other parameters,
such as $\dL$, and the errors estimated from the SNR alone by
equations~(\ref{eq:ddL(M,z)}) and (\ref{eq:dOmega(M,z)}) become less
accurate. In this regime, a better approximation to the scaling
of the errors is $ \propto (\Delta T / 1~{\rm yr})^{-1/2}\times
SNR^{-1}$, where $\Delta T \leq 1$~yr is the time elapsed from the
moment the binary appears at the low--frequency wall to the moment it
reaches the ISCO (Hughes 2005, private communication; see \cite{hh05}
for a more detailed treatment and discussion).

\section{Localizing the Counterparts}
\label{sec:DLerrors}

We next consider how to use the three--dimensional spatial
localization of the SMBH merger event by {\it LISA}.  The solid
angle error box directly yields the two--dimensional angular
position error on the sky, in which any EM counterpart will be
located. However, an additional step is necessary to convert the
luminosity distance, $\dL$,
measured by {\it LISA} into a redshift, $z$, which is the relevant
observable for the EM counterpart.  With a particular choice of
cosmological parameters $p_i=(H_0,\Omega_M,\Omega_{DE},w)$, we can
directly convert a redshift to a luminosity distance (see
eq.~\ref{eq:dL(z,p)} below), and vice--versa.  One may then envision
the following strategy: given the precision with which $p_i$ are
known from other observations, one can estimate the redshift, and
restrict the search for counterparts within the redshift shell
corresponding to the {\it LISA}-measured luminosity distance,
$\dL$. If a counterpart is
uniquely identified within this redshift shell, and its redshift can
be determined precisely, then one could hope for an improved
measurement of the $\dL(z)$ relation, and hence a refined
determination of the cosmological parameters.

The first step in this exercise is to determine the expected
redshift of the source. Apart from the errors on the cosmological
parameters and on the measurement of $\dL$, the peculiar velocity
of the source relative to
the Hubble flow, and its magnification due to weak gravitational
lensing by inhomogeneities in the distribution of mass along the
line of sight, introduce two additional sources of errors. In
summary, the redshift uncertainty will thus include a combination of
uncertainties from (i) the {\it LISA} luminosity distance, (ii) the
cosmological parameters, (iii) peculiar velocities, and (iv) weak
gravitational lensing magnification.

\cite{hou02} made a simplified estimate of the redshift error, without
peculiar velocities or gravitational lensing distortions, assuming a
flat cosmology with a cosmological constant (assuming $\Omega_M\equiv
1-\Omega_\Lambda$ and $w\equiv -1$), and ignoring correlations with
other cosmological parameters. Here, we extend that study by using a
general form of dark energy (relaxing the $w$ prior), by taking into
account the various parameter correlations, and by including errors
due to peculiar velocities and gravitational lensing.

To begin, we recall the luminosity distance to a source at a fixed
comoving coordinate in a smooth Friedmann universe,
\begin{equation}\label{eq:dL(z,p)}
\dL(z,p_i) = (1+z)c\int_0^z \frac{\D z'}{H(z',p_i)},
\end{equation}
where
\begin{equation}
H(z,p_i) = H_0 \sqrt{\Omega_M(1+z)^3 +
\Omega_{DE}(1+z)^{3(1+w)}}.
\end{equation}
We ignore spatial curvature and set $\Omega_k=0$, in line with
previous studies and as suggested by recent {\it WMAP} data.  For a
source with a small but non--zero radial peculiar velocity, $v$,
equation~(\ref{eq:dL(z,p)}) is modified, and the luminosity distance
is given by
\begin{equation}\label{eq:dL(z,p,v)}
\dL(z,p_i,v) = \dL[z_v,p_i,0],
\end{equation}
where $\Delta z\equiv z_v-z=(1+z)v/c$ is the additional apparent
redshift due to the peculiar motion.  In an inhomogeneous universe,
sources along different lines of sight can suffer different amounts
of gravitational lensing magnification $\mu$. If $\mu$ denotes the
magnification of the signal power, then the GW amplitude, and
thus the inferred value of $\dL^{-1}$, scales as $\mu^{1/2}$.  For a line of sight that
suffers a magnification $\mu$, equation~(\ref{eq:dL(z,p,v)}) is
modified as
\begin{equation}\label{eq:dL(z,p,v,mu)}
\dL(z,p_i,v,\mu) = \frac{1}{\sqrt{\mu}}\;\dL[z_v,p_i,0,0].
\end{equation}
In the limit of weak lensing $\mu^{1/2}\simeq 1 + \kappa$, where
$\kappa \ll 1$ denotes the weak lensing convergence (see below).

Taking the variation in both sides of equation~(\ref{eq:dL(z,p,v,mu)}),
we obtain
\begin{equation}\label{eq:delta dL}
\delta \dL = -\dL \kappa +\frac{\partial d_{\rm L}}{\partial z}\delta
z + \frac{\partial \dL}{\partial v} v + \sum_{i}\frac{\partial
\dL}{\partial p_i}\delta p_i,
\end{equation}
Solving for $\delta z$, taking the square and the expectation value of
both sides, and using the fact that the {\it LISA} measurement, the
peculiar velocities, and the cosmological uncertainties are
independent, i.e. $\langle \delta p_i\delta\dL\rangle=\langle
v\,\delta\dL\rangle =\langle \kappa\,\delta\dL\rangle=\langle v\,
\delta p_i\rangle=\langle \kappa \delta p_i\rangle= \langle \kappa v
\rangle =0$ for all $i$, we find
\begin{equation}\label{eq:delta_z^2}
\langle\delta z^2\rangle = \left(\frac{\partial \dL}{\partial
z}\right)^{-2}\left(
 \langle\delta d_{\rm L,LISA}^2\rangle +
 \langle\delta d_{\rm L,cosm}^2\rangle +
 \langle\delta d_{\rm L,pec}^2\rangle +
 \langle\delta d_{\rm L,wl}^2\rangle\right)
\end{equation}
or equivalently,
\begin{equation}\label{eq:delta_z}
\langle\delta z^2\rangle =
 \langle\delta z_{\rm LISA}^2\rangle +
 \langle\delta z_{\rm cosm}^2\rangle +
 \langle\delta z_{\rm pec}^2\rangle +
 \langle\delta d_{\rm wl}^2\rangle,
\end{equation}
where the notation was introduced to distinguish the intrinsic {\it
LISA} measurement error, $\delta d_{\rm L,LISA}$, from the error
resulting from cosmological parameters $\langle\delta d_{\rm
L,cosm}^2\rangle$, peculiar velocities $\langle\delta d_{\rm
L,v}^2\rangle$, and weak lensing magnification $\langle\delta d_{\rm
L,wl}^2\rangle$.  We now discuss each of these terms, whose forms and
magnitudes follow directly from equation~(\ref{eq:dL(z,p,v,mu)}).

\subsection{Cosmological uncertainties}

The cosmological term in equation~(\ref{eq:delta_z^2}) is
\begin{equation}\label{eq:delta d_Lc}
\langle\delta d_{\rm L,cosm}^2\rangle =\sum\limits_{i,j}\frac{\partial
\dL}{\partial p_i}\frac{\partial \dL}{\partial p_j}\langle\delta
p_i\,\delta p_j\rangle,
\end{equation}
where $\dL$ and its derivatives are to be evaluated using
equation~(\ref{eq:dL(z,p)}), at the fiducial values of the
cosmological parameters $p_i$, and for $v=\kappa=0$.  In order to
place {\it LISA} in the context of other experiments planned in the
next decade, we compute $\langle\delta d_{\rm L,cosm}^2\rangle$ from
the covariance matrices $\langle\delta p_i\,\delta p_j\rangle$
expected from two future cosmological probes: {\it
Planck}\footnote{See www.rssd.esa.int/index.php?project=PLANCK}
(assumed to have measured the temperature and polarization
anisotropies of the cosmic microwave background), and the Large
Synoptic Survey Telescope, LSST\footnote{See www.lsst.org} (assumed
to have measured the power spectrum and redshift distribution of
$\sim 100,000$ galaxy clusters). We adopt the forecasts for the
Fisher matrices for these experiments by \cite{wkhm04}.  Their
analysis assumes a flat background universe with 6 free parameters
$(\Omega_{DE}, \omega_M, w, \omega_b, n_s, \sigma_8)$ where
$\omega_M\equiv \Omega_M h^2\equiv(1-\Omega_{DE})h^2$ defines the
Hubble parameter.  Note that the luminosity distance does not
explicitly depend on $n_s$ and $\sigma_8$, but they are included
here, since they couple to the other four parameters as determined
by {\it Planck} and LSST (and hence increase the uncertainties on
these other four parameters). The fiducial parameters are
$(0.73,0.14,-1,0.024,1,0.9)$, respectively (consistent with {\it
WMAP}; \citealt{spergel03}).  Since the two observations are
independent, we simply sum up the two individual Fisher matrices;
the covariance matrix, $\langle\delta p_i\,\delta p_j\rangle$, is
obtained by taking the inverse of the Fisher matrix. In order to
substitute in equation~(\ref{eq:delta d_Lc}), it is necessary either
to revert to the original cosmological parameters
$(H_0,\Omega_{DE},\Omega_M,w)$ in the correlation matrix, by
performing an orthogonal transformation in the parameter
space\footnote{The parameters $\delta \Omega_M$ and $\delta
\Omega_{DE}$ will be fully anti-correlated, because of the
assumption of flatness.}, or to simply write
$\dL(z,p_i)$ of
equation~(\ref{eq:dL(z,p)}) in terms of the parameters
$(\Omega_{DE}, \omega_m, w)$ and evaluate the partial derivatives in
equation~(\ref{eq:delta d_Lc}) as a function of these parameters.
Following either approach, we find $\langle\delta d_{\rm
L,cosm}^2\rangle^{1/2}/d_{\rm L} = 1.7\times 10^{-3}$ for $z=1$. For
comparison, we performed the same analysis using the Fisher matrices
of WMAP (using only temperature anisotropies) and SDSS (using the
power spectrum of the luminous red galaxies, together with the
galaxies in the main SDSS survey, and following \citealt{hh03} for
redshift binning, mass limit, and sky coverage).  We find
$\langle\delta d_{\rm L,cosm}^2\rangle^{1/2}/d_{\rm L} = 1.1\times
10^{-2}$ for $z=1$ in that case.

\begin{figure}
\centering\mbox{\includegraphics[width=8.5cm]{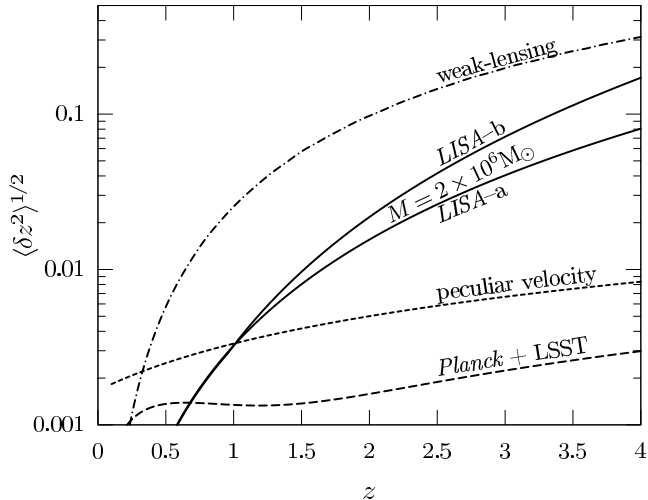}}
\caption[]{\label{fig:dz}
 Errors on the inferred redshift of an electromagnetic counterpart
 to a {\it LISA} coalescence event, for $m_1=m_2=10^6 M_\odot$. The
 intrinsic {\it LISA} error on the luminosity distance, $d_L$, is
 shown as a solid line for the two scalings $\delta \dL/\dL \propto
 SNR^{-1}$ (''{\it LISA}--a'') and $\delta \dL/\dL \propto \Delta
 T^{-1/2}\times SNR^{-1}$ (''{\it LISA}--b''). Errors due to the
 peculiar velocity of the source (for $v=500~{\rm km~s^{-1}}$;
 short--dashed line), uncertainties on the background cosmology
 (long--dashed line), and weak lensing magnification (dash--dotted
 line) are also shown (see text for details). }
\end{figure}

The luminosity distance error corresponds to a redshift error
according to equation~(\ref{eq:delta_z^2}).  Note that
equation~(\ref{eq:delta d_Lc}) depends on the fiducial redshift
through the $\dL$ derivatives. This dependence of the redshift error
is shown as a long--dashed curve in Figure~\ref{fig:dz}, along with
other sources of redshift errors.  The luminosity distance at
$z\approx 1000$ is measured very accurately by {\it Planck}, and its
evolution is essentially unaffected by the cosmological parameters
down to dark-energy domination at $z\lsim 2$. The figure shows that,
as a result, the relative cosmological error $\langle\delta z_{\rm
cosm}^2\rangle^{1/2}/z$ reaches a constant value beyond $z\gsim 2$. The figure also
shows that the cosmology error becomes smaller than the typical LISA
uncertainty at $z\gsim 0.7$.  We find that, even for our best--case
{\it LISA} events, the cosmology error becomes sub--dominant at
$z\gsim 1$.

\subsection{Peculiar velocities}

Equations~(\ref{eq:dL(z,p,v)}), (\ref{eq:delta dL}), and
(\ref{eq:delta_z^2}) yield, to first non-vanishing order in $v/c$, the
uncertainty due to peculiar velocities, as a simple function of
the r.m.s. peculiar velocity of the GW sources,
\begin{equation}\label{eq:delta d_Lv}
\langle\delta d_{\rm L,pec}^2\rangle/\dL^2 = \left[ 1 + \frac{c
(1+z)^2}{H(z)\dL}\right]^2  \frac{\langle v^2\rangle}{c^2}.
\end{equation}

We are assuming here that GW events correspond to luminous quasars,
and unfortunately, the r.m.s. peculiar velocity of high-redshift
quasars is not known empirically. We therefore employ theoretical
predictions for the peculiar velocities of (quasar--host) galaxies
within dark matter haloes. According to numerical simulations (e.g.
\citealt{co00}), typical values are $\sim 500~{\rm km~s^{-1}}$, with a
tail extending to $\sim 1000~{\rm km~s^{-1}}$.  As an approximation,
we assume that the scaling with redshift follows the linear growth in
the amplitude of density perturbations, multiplied by the linear bias
of the halos. Under this assumption, we find that for fixed halo mass,
the typical peculiar velocity evolves very little from $z\sim 0$ to
$z\sim 1$. This conclusion is consistent with the semi-analytic model
of \cite{hkysj03}, which shows essentially no evolution (only a modest
decrease from $z\sim 0$ to $z \sim 0.8$).  Therefore, we substitute $v
\sim 500~{\rm km~s^{-1}}$ in equation~(\ref{eq:delta d_Lv}) at $z\sim
1$, yielding an error of $\langle\delta d_{\rm
L,pec}^2\rangle^{1/2}/\dL=4.1\times 10^{-3}$. The corresponding
redshift r.m.s. error contribution (equation~[\ref{eq:delta_z}]) for
$v \sim 500~{\rm km~s^{-1}}$ is shown as a function of redshift in
Figure~\ref{fig:dz} (short--dashed curve). The figure shows that the
peculiar velocity error is lower than the typical {\it LISA} error for
$z\gsim 1$.

Note that the r.m.s. peculiar velocity equals $\sim 500~{\rm
km~s^{-1}}$ in the most typical cases, but its exact value depends on
the specific mass of the halo, $M_{\rm halo}$, embedding the quasar
\citep{hkysj03}, smaller $M_{\rm halo}$ implying lower velocities. For
a specific source, $M_{\rm halo}$ can be estimated directly, using the
number of galaxies that cluster around the identified quasar
\citep{kh02}. The r.m.s. peculiar velocity error may then be estimated
for this particular source, and may be somewhat lower/higher than the
value shown in Figure~\ref{fig:dz}, depending on whether the source is
located in a galaxy--poor/galaxy--rich environment.

\subsection{Weak Gravitational Lensing}

The gravitational lensing term in equation~(\ref{eq:delta_z^2}), in the
weak-lensing limit, is given by
\begin{equation}\label{eq:delta d_Lwl}
\langle\delta d_{\rm L,wl}^2\rangle/\dL^2 = \langle\kappa^2\rangle,
\end{equation}
where $\kappa$ denotes the r.m.s. effective convergence (e.g.,
\citealt{wh00}).  While the mean magnification is well approximated by
$\langle\kappa\rangle=0$ or $\langle\mu\rangle=1$, the magnification
distribution has a substantial width.  The full distribution is given
by \cite{whm02}, and its variance reaches $\sim 12\%$ for sources at
$z=2$ \citep{dhcf03}.  Here we use equation~(6) in \cite{wh00} to
compute the variance in effective convergence,
$\langle\kappa^2\rangle$, for point sources.  This quantity is given
by an integral over the matter power spectrum, and receives a
contribution from small, non--linear scales. We employ the HALOFIT
routine of \cite{smith03}, and set the input cosmological parameters
according to our convention (see \S~\ref{sec:Introduction}). This
routine encodes an accurate fitting formula for the matter power
spectrum extending into the nonlinear regime. For $z=1$, we find
$\langle\kappa^2\rangle^{1/2}= 3.1\%$.

Weak lensing errors can, in principle, be improved by ``corrective
lenses'' (\cite{dhcf03}), i.e. background galaxy shear maps, and using
the cross-correlation between the magnification of a point source and
the shear map smoothed on larger--scales.  However, \cite{dhcf03}
found that the magnification errors can be improved by only a small
amount, less then 20\% relative to the uncorrected errors for a source
at $z=2$.

A different approach for reducing weak lensing magnification
uncertainty would be to directly measure the inhomogeneities in the
mass distribution along the line of sight.  If this distribution could
be directly probed down to a scale of $k_{\min}$, then the
contributions to $\kappa$ from all scales down to $k_{\min}$ could be
directly subtracted from the uncertainty on the magnification.  If the
counterpart is indeed a quasar, then the line-of-sight density
distribution could, in principle, be probed by studying its Lyman
$\alpha$ absorption spectrum, as well as deep surveys of galaxies and
clusters in the foreground and near the line of sight.  At low
redshifts ($z\sim 0.5$) which contribute significantly to the lensing
magnification, the X-ray absorption forest \citep{fc00,pl98} could
provide additional information on the density fluctuations.

\begin{figure}
\centering\mbox{\includegraphics[width=8.5cm]{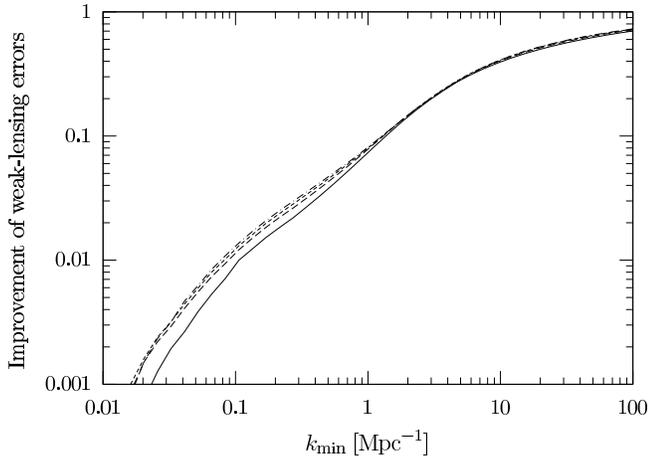}}
\caption[]{\label{fig:wlkmin} The relative improvement in the weak
lensing magnification--induced redshift uncertainty to a background
source.  We show the fractional improvement that can be achieved by
perfectly measuring density inhomogeneities down to a fixed scale.
The $x$-axis shows the minimum wave number that is unmapped (larger
scale fluctuations are assumed to be perfectly known along the line of
sight to the source). Various curves correspond to sources at $z=1$,
2, 3, and 4 (from bottom to top).}
\end{figure}

We leave a detailed assessment of the amount of correction that
could be feasible to future investigations.  However, to estimate
the ``target'' scale at which the fluctuations would need to be
measured in order to be useful as a lensing correction, we make the
(unrealistically optimistic) simplifying assumption that the matter
fluctuations above a certain length scale, and the corresponding
contribution to lensing magnification, have been perfectly
determined. The fluctuations on smaller scales, with $k>k_{\min}$,
then determine the only remaining weak-lensing uncertainty.
Therefore, we truncate the integral over the wavenumber (see eq.~[1]
in \citealt{dhcf03}) at $k_{\min}$. The resulting fractional
improvement in $\langle\kappa^2\rangle^{1/2}$ (which equals $\delta
d_{\rm L, wl}/\dL$) is plotted in Figure~\ref{fig:wlkmin}. The
improvement is about 8\%, 20\%, and 40\% for $k_{\min}=1$, 3, and
$10\Mpc^{-1}$, respectively - i.e., about half of the weak lensing
uncertainty is from length--scales $\lsim 2\pi/k_{\min}\sim 0.6$
Mpc. The Lyman $\alpha$ transmission spectra of SDSS quasars have
been used to determine the power spectrum down to scales of $k \sim
3 \Mpc^{-1}$, implying that the spectral resolution would need to be
improved by a factor of $\sim 3$ to allow significant improvements.

The uncorrected weak-lensing redshift uncertainty $\delta z_{\rm
wl}\equiv(\partial \dL/\partial z)^{-1}\langle\delta d_{\rm
L,wl}^2\rangle^{1/2}$ is shown in Figure~\ref{fig:dz}. The plot
clearly shows that weak lensing errors typically exceed 1\% and
dominate the other redshift errors for $z>0.5$.

\subsection{The Size and Orientation of the Error Volume}
\label{sec:volumeshape}

In Figure~\ref{fig:dz}, we compare the contributions from four
different sources of redshift errors: $\langle \delta z_{\rm
LISA}^2\rangle^{1/2}$, $\langle\delta z_{\rm pec}^2\rangle^{1/2}$,
$\langle\delta z_{\rm cosm}^2\rangle^{1/2}$, and $\langle\delta
z_{\rm wl}^2\rangle^{1/2}$ for SMBH masses $m_1=m_2=10^6 {\rm
M_{\odot}}$. Weak lensing errors are dominant for $5\gsim z\gsim
0.5$. The {\it LISA} redshift error is shown for both scalings
$\delta \dL/\dL \propto SNR^{-1}$ (labeled ''{\it LISA}--a'') and
$\delta \dL/\dL \propto (\Delta T /1~{\rm yr})^{-1/2}\times
SNR^{-1}$ (labelled ''{\it LISA}--b''). These curves separate for
$M_z > 3.92\times 10^6 {\rm M_\odot}$, when the maximum observation
time $\Delta T$ decreases under $1$~yr. The other three sources of
errors are within a factor of two of one another at $z=1$, for
typical {\it LISA} sources and peculiar velocities.

The total redshift uncertainty then follows from summing the
uncertainties in quadrature (eq.~[\ref{eq:delta_z}]). The result for
$m_1=m_2=10^6~{\rm M_\odot}$ and $z=1$ is $\delta z=2.57\%$, 2.59\%,
and 3.03\% in the cases we labeled as best, typical, and worst,
respectively. Note that we have not explicitly added an error due to
the instrumental resolution of a spectrograph, because this is often
much better than this value (e.g. for
SDSS\footnote{http://cas.sdss.org/dr3/en/tools/search/sql.asp}, the
redshift resolution is between $10^{-3}$ and $10^{-4}$).

We next use these redshift uncertainties to derive the
three--dimensional error volume, $\delta V_{\rm tot}$, in which the EM
counterparts need to be identified.  The comoving volume corresponding
to the above redshift uncertainties, combined with the solid--angle
uncertainty $\langle\delta\Omega^2\rangle^{1/2}$, is
\begin{equation}\label{eq:dV}
\delta V_{\rm tot} =
 \frac{\partial^2 V}{\partial z\partial\Omega}
 \langle \delta z^2 \rangle^{1/2}
 \langle\delta\Omega^2\rangle^{1/2}.
\end{equation}
Substituting the solid--angle errors from Table~\ref{tab:LISA}, we
find $\delta V_{\rm tot}= 2.1\times 10^3$, $6.3\times10^{4}$, and
$7.4\times10^{5} \Mpc^3$ for the same three cases.

Equation~(\ref{eq:dV}) should be regarded as a simple estimate of
the volume that needs to be searched.  The 2D uncertainty
$\langle\delta\Omega^2\rangle^{1/2}$ was obtained by a Fisher
analysis, and represents the solid area of a 2D error ellipse. This
equation then gives the volume of a cylinder with a height of
$\langle \delta z^2 \rangle^{1/2}$.  The arguably more appropriate
volume of an ellipsoid whose third semi--axis is $\langle \delta z^2
\rangle^{1/2}$ would be a factor of 4/3 larger. On the other hand,
equation~(\ref{eq:dV}) also naively assumes that $\delta z$ and
$\delta \Omega$ are uncorrelated -- i.e., it describes an error
ellipsoid whose $z$ axis is oriented along the line of sight. The
angular position is estimated by {\it LISA} from the GW signal
alone, and is therefore indeed uncorrelated with the radial
uncertainties due to cosmology, lensing, or peculiar velocities.
However, the luminosity distance estimate from {\it LISA} itself is
strongly correlated with its angular positioning \citep{hou02},
which results in an error ellipse that is 'tilted' relative to the
line of sight, and has a smaller overall volume than the simple
orthogonal product in equation~(\ref{eq:dV}) would imply.

We have utilized the correlation matrices for the {\it LISA}
distance and angle estimates given in Table 1 and 2 of \cite{hou02}
for $m_1=m_2=10^5{\rm M_{\odot}}$ at $z=1$ and $m_1=m_2=10^4{\rm
M_{\odot}}$ at $z=7$, to estimate the reduction in the total error
volume due to these correlations.  Note that this analysis applies
only to the {\it LISA} uncertainties. Among the 11 free parameters
in \cite{hou02}, the 3D error volume is determined by the parameters
related to the spherical coordinates of the sources: $\ln
\dL$, $\mu_S=\cos\theta_S$,
and $\phi_S$.  The error volume $\delta V_{\rm tot}$ quoted in
equation~(\ref{eq:dV}) above corresponds to (3/4th of) the volume of
an ellipsoid whose semi-axes are the {\it marginalized} errors in
spherical coordinates; the true error volume $\delta V_{\rm ell}$ is
that of the ellipsoid described by the full covariance matrix.  The
ratio $\delta V_{\rm ell}/\delta V_{\rm tot}$ depends on the actual
position angle ($\theta_S,\phi_S$) of the source; averaging over all
angles, we find $\langle \delta V_{\rm ell}/\delta V_{\rm tot}
\rangle = 0.31$ (0.20) for masses $m_1=m_2=10^5 {\rm M_\odot}$
($10^4$) at redshift $z=1$ (7).

If {\it LISA} errors dominated the total redshift uncertainty $\langle
\delta z^2 \rangle^{1/2}$, this implies that the correlations could
reduce the mean number of counterparts by a factor of 3--4.  However,
as discussed above, the total uncertainty is likely going to be
dominated by weak lensing errors; hence the inclusion of the
correlations would reduce the final error volume only by a small
factor ($\sim 15\%$).  We make no use of this reduction in the results
we quote below.

\section{Quasar counterparts}
\label{sec:QSOLF}

To estimate the typical number of quasar counterparts to a specific
SMBH merger event, we combine the size, $\delta V_{\rm co}$, of the
comoving {\it LISA} error box with the space density of quasars, by
integrating over the quasar luminosity function, $\phi(L,z)$:
\begin{equation}\label{eq:number of cp}
N = b \; \delta V_{\rm co}\int_{L_{\min}}^{L_{\max}} \D L\;
\phi(L,z),
\end{equation}
where $b$ accounts for the bias due to the clustering of quasars,
and $L_{\min}$ and $L_{\max}$ are the minimum and maximum quasar
luminosities which could be associated to the specific SMBH merger
event. We use $L_{\min}=0.1 L_{\rm Edd}$ and $L_{\max}=2 L_{\rm
Edd}$, where $L_{\rm Edd}$ denotes the Eddington luminosity for the
total BH mass. Motivations behind this particular near-Eddington
choice are further discussed below.

\subsection{Luminosity Function of Quasars}

We adopt the standard empirical double power--law fit to the quasar
luminosity function of the combined quasar samples from the
Two-Degree Field (2dF) and Sloan Digital Sky Survey, with pure
luminosity evolution valid for $z < 2.1$ \citep{ric05,croom05}.
Unfortunately, these surveys extend only to relatively bright
magnitudes ($B\lsim 21$), corresponding to the Eddington luminosity
of BHs with mass $M\gsim 3 \times 10^7~{\rm M_\odot}$ for $z=1$,
which is above {\it LISA}'s optimal mass--range. In order to
estimate the number of quasar counterparts for lower BH masses, we
extrapolate the luminosity function to the faint end.

A limitation of this luminosity function is the simple quadratic
fitting formula for the evolution \citep{ric05,croom05}, which is
only valid for redshifts $z<2.1$. To avoid these difficulties
\cite{mhr99} (hereafter MHR) proposed a more complicated empirical
fitting formula with three adjustable parameters $z_s$, $\zeta$, and
$\xi$.  These parameters were estimated using high-$z$ quasars, and
were improved to include the high-redshift SDSS sample and
weak-lensing effects \citep{wl02}. We obtain a luminosity function
that is more precise at lower redshifts and is concordant with the
redshift scaling by fitting the MHR model to the \cite{ric05}
luminosity function in the interval $0.5<z<2.1$ and keeping the
high--redshift asymptote at the \cite{wl02} value.  The result is
$L^{*}=5.06\times 10^{10} {\rm L_{\odot}}$, $z_s=1.66$, $\zeta=2.6$,
and $\xi=2.8$. Other parameters in the luminosity function that are
independent of the evolution are adopted from \cite{ric05}:
$\beta_{\rm l}=1.45$,
$\beta_{\rm h}=3.31$, and $\Phi^{*}_{\rm L}=1.99 \times
10^{-6}\Mpc^{-3}$.

\subsection{Clustering of Quasars}
The bias, $b$, in equation~(\ref{eq:number of cp}) describes the
enhancement in the number of quasars around a specific quasar being
the potential counterpart, relative to the value for a homogeneous
distribution.  The clustering depends only weakly on quasar
luminosity \citep{lidz05,as05}. We use the observed autocorrelation
function of quasars from the 2dF survey \citep{croom05},
$\xi(s)=(s/s_0)^{\gamma}$ for $s>0.1 h^{-1}$Mpc (assuming no quasars
with a smaller separation), where $s_0=5.48 h^{-1}$Mpc (5.55) and
$\gamma=1.20$ (1.63) for $s<25\Mpc$ (>25Mpc), respectively.  The
bias is given by the average value of $\langle 1+\xi(t)\rangle$ over
the comoving error box. Assuming that the error box is a cylinder,
with height $\delta y = c\delta z/H(z)$ (which is the comoving
distance along the line of sight corresponding to the redshift
error) and radius $\delta r = \sqrt{\delta V_{\rm co}/(\pi\delta
y)}$ (corresponding to the angular uncertainty of {\it LISA}):
\begin{equation}
b = \frac{\int_0^{\delta r} 2\pi r \D r \int_0^{\delta y} \D y\;
(1+\xi(\sqrt{y^2+r^2}))}{\int_0^{\delta x} 2\pi r \D r\int_0^{\delta
y} \D y}.
\end{equation}
We find $b=1.50$, 1.23, and 1.07, for our best, typical, and worst
cases, respectively. The corresponding comoving cylinder heights are
$\delta y = 62.5$, 62.9, and 73.8 Mpc, and the cylinder radii are
$\delta r = 3.3$, 17.8 and 56.4 Mpc.

\begin{deluxetable}{lllrl} \tablecolumns{5}
\tablewidth{0pt} \tablecaption{\label{tab:surveys}Survey
characteristics} \tablehead{\colhead{Survey} & \colhead{$M_{\min}$} &
\colhead{Limiting mag} & \colhead{Sky cov.} & \colhead{Observing}\\
\colhead{} & \colhead{$(z=1)\;{\rm M_{\odot}}$} & \colhead{} &
\colhead{$\deg^2$} & \colhead{}}
 \startdata
 2dF\tablenotemark{a} (2QZ,6QZ)& $3\times10^{7}$ & $b_{J}<20.85$ & $750$ & 1997-2002 \\
 SDSS\tablenotemark{b} (LRG) & $2\times10^{8}$ & $i<19.1$ & $7\times 10^3$ & 1998-2005 \\
 Deep2\tablenotemark{c} & $5\times10^{5}$ & $R<24.5$ & $4$ & 2002-04 \\
 AGES\tablenotemark{d} & $1\times10^{7}$ & $R<21.5$ & $9$ & 2004-06 \\
 DES\tablenotemark{e} & $7\times10^{5}$ & $AB<24.7$ & $5\times 10^3$ & 2009-13 \\
 LSST\tablenotemark{f} & $2\times10^{5}$ & $AB<26.5$ & $1.8\times 10^4$ & 2012 \enddata
\tablenotetext{a}{Two-Degree Field, see http:/www.2dfquasar.org/}
\tablenotetext{b}{Sloan Digital Sky Survey, see http:/www.sdss.org/}
\tablenotetext{c}{See http:/deep.berkeley.edu/}
\tablenotetext{d}{AGN and Galaxy Evolution Survey, covers radio, IR,
optical, and X-ray bands, see
http:/cmb.as.arizona.edu/$\sim$eisenste/AGES/} \tablenotetext{e}{Dark Energy Survey, see
http:/cosmology.astro.uiuc.edu/DES/ and http:/decam.gnal.gov/ }
\tablenotetext{f}{Large Synoptic Survey Telescope, see
http:/www.lsst.org/}
\end{deluxetable}

\section{Results}
\label{sec:results}

\begin{figure}
\centering{
 \mbox{\includegraphics[width=8.5cm]{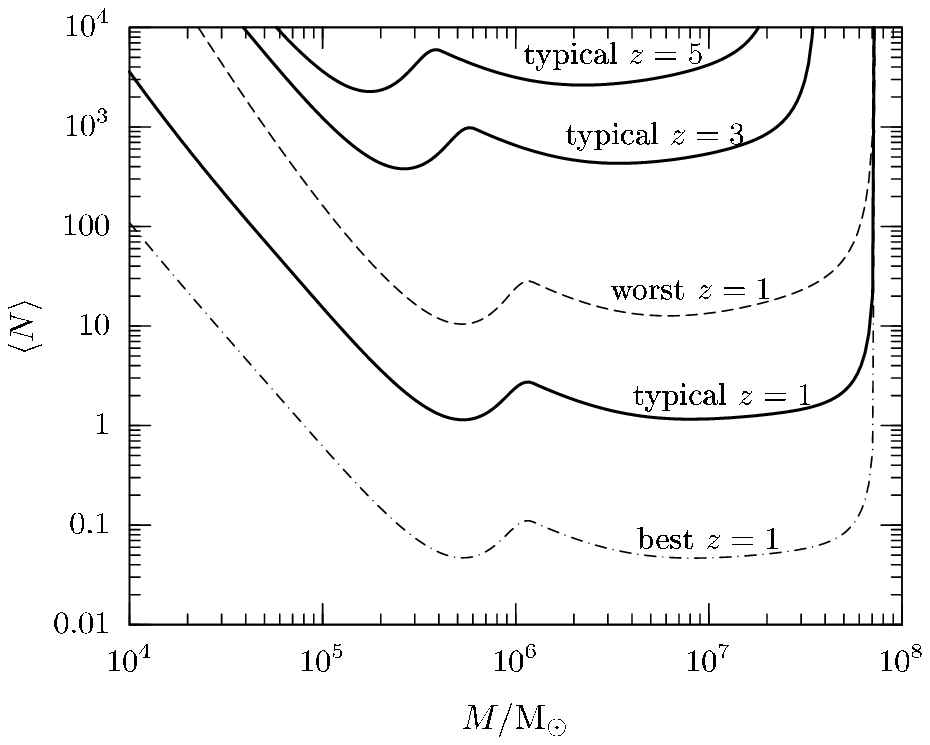}}\\
 \mbox{\includegraphics[width=8.5cm]{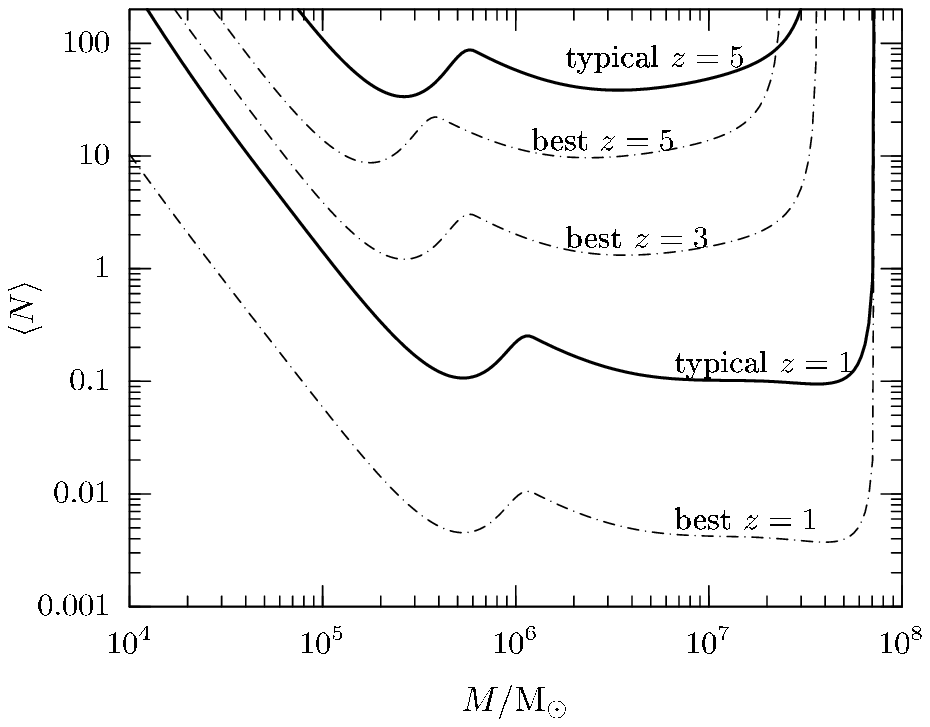}}\\
 \mbox{\includegraphics[width=8.5cm]{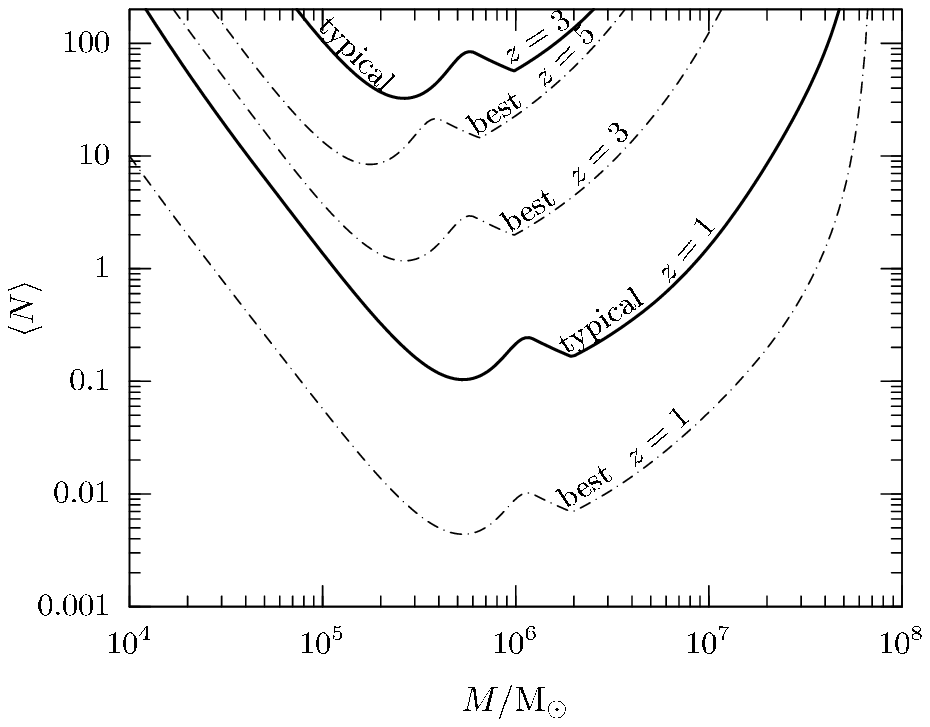}}}
\caption[]{\label{fig:cp1} The average number of quasars in the
three--dimensional {\it LISA} error volume. The thin dash--dotted,
thick solid, and thin dashed curves trace the ''best'', ''typical'',
and ''worst'' cases, respectively. The low--frequency noise wall for
{\it LISA} is assumed to be at $f_{\min}=0.03\mHz$.
 {\it Top}: Using raw data without any weak-lensing corrections and
    the counterpart's luminosity is assumed to be $0.1<L/L_{\rm Edd}<2$.
 {\it Middle}: The weak-lensing errors are corrected by 20\%, and the
    luminosity is assumed to be $0.7<L/L_{\rm Edd}<1.3$.
 {\it Bottom}: Same as the middle panel, assuming that {\rm LISA}
    uncertainties scale as
    $\delta\dL/\dL\propto \Delta T^{-1/2}\times SNR^{-1}$ and
    $\partial\Omega\propto \Delta T^{-1}\times SNR^{-2}$.
In most cases at $z\lsim 1$, a unique quasar counterpart may be
identifiable.}
\end{figure}

The main results of this paper are presented in Figure~\ref{fig:cp1},
which shows $\langle N\rangle$, the average number of counterparts
within the 3D {\it LISA} error volume, for various total masses $M$
and redshifts $z$.  Recall that we have assumed that the GW event is
always accompanied by quasar activity.  According to our definition,
$\langle N\rangle$ then corresponds to the mean number of quasars that
would be found in {\it LISA}'s error box, {\it in addition} to the
quasar actually associated with the GW source.  A straightforward
identification of a unique counterpart therefore requires that there
be no additional quasars in the error volume and that the EM
observation sensitivity goes below the actual counterpart luminosity
(so that the presence of fainter quasars can be ruled out). A simple
criterion for a reasonable chance not to have any additional
counterparts candidates is
$\langle N\rangle <1$.\footnote{One could explicitly consider the probability
distribution for $N$. For example, for a Poisson distribution, the
probability for $N=0$ would be 50/90 percent for $\langle N\rangle=0.7/0.1$.}
We find that this simple condition is
satisfied in the case of a ''typical'' event at $z=1$ with total
masses $\sim 4\times 10^5~{\rm M_\odot}$ or $\sim 8\times 10^6~{\rm
M_\odot}$. At higher redshifts, the average number of potential
counterparts will be much larger, due mostly to the increasing weak
lensing errors. At $z \gsim 3$, even the best case events will typically
have at least one additional quasar in their error box. On the other
hand, the increase from $z=3$ to $z=5$ in the number of quasars
located in the error-box is partly mitigated by the drop in the
abundance of quasars at $z \gsim 3$.  The three panels of
Figure~\ref{fig:cp1} display results for various presumptions of
uncertainties. The top panel uses raw, uncorrected weak-lensing errors
and the counterpart luminosity is allowed to vary in a broad range
$0.1<L/L_{\rm Edd}<2$. The middle panel accounts for a 20\% reduction
of weak-lensing errors, and the luminosity is restricted to
$0.7<L/L_{\rm Edd}<1.3$. The bottom panel shows results for the more
conservative scaling of {\it LISA} errors, $\delta\dL/\dL\propto
\Delta T^{-1/2}\times SNR^{-1}$ and $\partial\Omega\propto \Delta
T^{-1}\times SNR^{-2}$ if $\Delta T<1\yr$. In all cases at $z\sim 1$,
a unique counterpart may be identifiable.

One may also interpret the number of counterparts we compute, $\langle
N\rangle$, together with the same simple criterion $\langle N\rangle <
1$, in a somewhat different way: as a means to {\it test} our hypothesis
that {\it LISA} events are accompanied by bright quasar activity.  If
our hypothesis is incorrect, then the number of quasars in {\it
LISA}'s error volume should be drawn from the random distribution of
quasars on the sky, unrelated to the {\it LISA} event (excluding the
correction due to correlations that we have included in our analysis;
although we found this correction to be relatively insignificant).  In
many configurations, we find $\langle N\rangle\ll 1$, implying a
significant probability that {\it no} bright quasars would be found in
{\it LISA}'s error volume in these cases.  If several {\it LISA}
events are indeed found with no such quasar counterparts, it would, by
itself, be an important new constraint on the process of binary black
hole coalescence.

The identification of counterparts could be aided by a combined
analysis of several GW events. Every successful identification yields
a very precise direct measurement for the $L/L_{\rm Edd}$ Eddington
ratio. Once a statistically significant set of Eddington ratios is
acquired, the empirical distribution can be mapped.
It is yet unclear whether SMBH mergers are expected to have
high-luminosity quasar counterparts.
Then, {\it in case} $L/L_{\rm Edd}$ is in a narrow range, this
information can be used to greatly constrain the a priori assumptions
on the counterpart's luminosity. As an example, in
Figure~\ref{fig:cp1} (middle panel) we consider
$L/L_{\rm Edd}=1\pm 0.3$, and find $\langle N\rangle$ to decrease
well under 1 in the typical case.

If cosmological uncertainties were to dominate the error budget on a
counterpart's redshift, a combined analysis could further improve the
robustness of the identification. Indeed, even if each GW event has,
by itself, several possible counterparts, each of these counterparts
would require a different set of cosmological parameters. As a result,
there should be, in general, only a single set of cosmological
parameters\footnote{This set, of course, will suffer from the usual
degeneracy along the surface of constant $\dL(z,\Omega_\Lambda,
\Omega_m,h,w)$ around the fiducial cosmological parameters.}  that
gives a consistent set of counterparts to all of the GW events
\citep{s86}.  The counterpart candidates contradicting this concordant
set can be discarded.  Unfortunately, the error budget is likely to be
dominated by the lensing magnification uncertainty. In this case,
having multiple events is going to be helpful only if a sufficient
number ($\gg 100$) of events are detected to map out the full
distribution of magnifications (such a large number of events is not
expected; see below).

\section{Discussion}
\label{sec:discussion}
\subsection{Search Strategy}

Different strategies for the search of an electromagnetic
counterpart to a {\it LISA} event can be envisioned. The simplest
one would be to search, in an existing survey catalog, for
candidates located in the 3D {\it LISA} error box. We list the
characteristics of several deep present and near-future surveys in
Table~\ref{tab:surveys}. These surveys cover various bands between
400 and 1100nm. In each case, we calculate the minimum mass of the
quasar corresponding to the magnitude limit of the survey, provided
that the quasar is shining at the Eddington luminosity.  For the
quasar spectrum we use \cite{mhr99} and follow \cite{madau95} to
account for the Lyman series line blanketing and photoelectric HI
absorption to obtain the apparent luminosity.  The portion of the
sky covered and scheduled operation times are also given.  The most
promising (deepest and largest sky coverage) among the planned
surveys is that from the LSST, scheduled for the same time as {\it
LISA}.

A second strategy would be to design a survey specifically aimed at
identifying the counterparts of well--localized {\it LISA} events,
by observing in detail the 2D angular error box provided by {\it
LISA} and looking for the expected counterparts. The time domain may
come to play an important role in the strategy of this type of
surveys. Indeed, for some of the events, one may have, from the {\it
LISA} data-stream, a reasonably good idea of the location on the sky
of the SMBH binary prior to coalescence. For instance, \cite{hh05}
estimate that this knowledge may be available with reasonable
accuracy about a day in advance. One would then be able to monitor
any unusual photometric variability associated with the violent SMBH
merger in this area somewhat before, during and after the
coalescence. Surveying the area long (i.e. months or years) after
the coalescence may also prove useful in discovering the counterpart
and/or monitoring the viscous evolution of any gas surrounding the
SMBH merger remnant \citep{mp05}. The time domain may thus greatly
facilitate the identification of a unique electromagnetic
counterpart to some {\it LISA} events, even in cases when many
counterpart candidates otherwise exist in the 3D error box.

\subsection{Uncertainties in the Analysis}

Throughout the analysis of {\it LISA} uncertainties, we focused
on equal mass binaries. With this simplifying assumption, it was possible
to derive approximate scalings as a function of total mass and relate
them to the calculations of \cite{vec04} for $m_1=m_2=10^6 M_{\odot}$.
We also find good agreement with the recent calculations of
\cite{hh05} for a variety of equal-mass combinations. For unequal
masses, the GW signal-power depends to leading order on the simple
combination ${\cal M} = \mu^{3/5}M^{2/5}$, allowing a straightforward
extension of our scaling arguments to more general mass ratios.

As mentioned above, our results depend on the {\it LISA} sensitivity
curve at low frequencies. In particular, the relationship between
the arrival frequency $f_{\rm a}(M_z,\Delta T)$, the final frequency
$f_{\isco}(M_z)$, and the {\it LISA} minimum frequency noise wall
$f_{\min}$ determines the maximum possible observation time for the
in-spiral. The exact value of $f_{\min}$ is currently assumed to lie
between 0.1 and 0.03mHz.  Figure~\ref{fig:cp1} assumes
$f_{\min}=0.03\mHz$.  In this case, at $z=1$, a $\Delta T=1\yr$
observation of the in-spiral phase is possible for $M<2\times 10^6
{\rm M_{\odot}}$ but for larger masses, the maximum possible
observation time becomes shorter than a year.  We have also computed
the number of quasars in the 3D {\it LISA} error box for the higher
value of $f_{\min}=0.1\mHz$. We find a significant increase in this
number at the high--mass end (i.e. for $M \sim 10^7 {\rm M_\odot}$),
but the results are essentially unaffected for lower mass SMBHs.

An important assumption in our analysis is the near-Eddington
luminosity of the quasar counterparts associated with {\it LISA}
events. While this assumption is difficult to justify from first
principles, it is the luminosity expected if SMBH accretion occurs
in a regime which is not supply-limited.  Observationally, the
Eddington ratio of quasars, $L/L_{\rm Edd}$, can be inferred,
leading to values from 0.1 to $\gsim 1$, with higher values at large
redshifts (e.g. \citealt{kas00,ves04,wu02}). This ratio has
been determined for a handful of lower mass SMBHs in AGNs and these
sources have been found to cluster around $L/L_{\rm Edd}\sim 1$ as
well \citep{gh04}. Recent work by \cite{kol05} and \cite{hop05}
both suggest $L/L_{\rm Edd}$ is typically around 1/3.

The restricted luminosity range $0.1<L/L_{\rm Edd}<2$ assumed in our
analysis effectively serves as a 4th (``brightness'') dimension,
complementing the three-dimensional geometrical error volume provided
by {\it LISA}.  According to Table~\ref{tab:LISA}, for a given event,
the mass estimate provided by {\it LISA} is typically very accurate.
Therefore, it is through the range of acceptable Eddington ratios that
the integration bounds in equation~(\ref{eq:number of cp})
change. Since the luminosity function decreases rapidly with $L$, the
integral is dominated by the lower bound and is largely independent of
$L_{\max}$. We chose $L_{\max}=2L_{\rm Edd}$ for concreteness. When
modifying the lower bound from $L_{\min}=0.1 L_{\rm Edd}$ to
$L_{\min}=1.0 L_{\rm Edd}$ or $L_{\min}=0.01 L_{\rm Edd}$, for
instance, we find that the number of quasars present in the 3D {\it
LISA} error box changes by a factor of 0.1 and 5,
respectively.\footnote{The change is explained by the asymptotic form
of the luminosity function $\Phi(L)\propto L^{-1.45}$ for low $L$.}
Obviously, our need to extrapolate the quasar luminosity function
below the minimum value constrained by the observations is an
additional important source of uncertainty in our analysis, but one
that will be addressed well by future generations of surveys
(Table~\ref{tab:surveys}).

In a recent study, \cite{hennawi05} have found a large number of
small--separation quasars, implying an order-of-magnitude increase in
the quasar auto-correlation on scales $\lsim 200 h^{-1}$ kpc. However,
the dimensions of the error volume as we estimated it is generally
much larger ($\sim3.3$ Mpc even in our best case). As a result,
the average number of counterparts is changed by a negligible factor
by this small--scale auto-correlation: $8.3\times10^{-3}$,
$3.7\times10^{-4}$, and $3.3\times10^{-5}$ in the best, typical, and
worst case {\it LISA} errors. Increasing the correlation by an order
of magnitude out to a scale of $500 h^{-1}$ kpc would still cause at
most a 6\% increase in the mean number of quasars in the best
case. \cite{hennawi05} do not find an increase in the correlation
beyond $\sim 500 h^{-1}$ kpc scales.  We note that our inability to
identify the correct counterpart from among any rare
ultra-small-separation candidates would not degrade cosmological
parameter estimations significantly. One may also argue that such
close--separation binary quasars would better represent a 'precursor'
stage in the evolution of the two black holes towards an eventual
coalescence. If so, one would not expect them to be associated
with GW events, and they may in fact be anti--correlated with such
events.

Throughout this paper, we have focused on optical quasars as
plausible counterparts to GW events. It would be interesting to
repeat our analysis using other types of ``electromagnetic objects''
that may be associated with SMBH coalescences. For example, even if
gas accretion leads to prodigious energy output in radiation, the
optical light of the quasar may be obscured by the intervening gas
and dust near the galactic nucleus.  In these cases, the GW events
may be more commonly associated with X--ray quasars (e.g.
\citealt{mp05}), or with ultraluminous infrared galaxies (e.g.
\citealt{thomp05}). As these sources are also relatively rare,
unique counterparts may be identifiable if such objects typically
accompany GW events.

\subsection{Implications: Black Hole Astrophysics}

A successful identification of a quasar counterpart to a {\it LISA}
event would provide powerful diagnostics on the physics of SMBH
gaseous accretion and the associated radiation. The masses and spins
of the two BHs before merger can be directly determined from the GW
signal, from which the mass and spin of the remnant BH follows at some
precision. In some cases (e.g. Hughes \& Menou 2005), the remnant BH
could also be observed by {\it LISA} during the post-merger ring-down
phase, which would constrain its mass and spin directly. The
orientation of the orbital plane of the BH binary before merger would
be measured as well. All of these parameters are generally unknown for
quasars detected only via traditional electromagnetic techniques.

The observation and monitoring of quasar counterparts to {\it LISA}
events may thus offer us some of the best laboratories for the study
of AGN physics. First, the Eddington ratio can be measured to high
accuracy (limited only by photometric errors and bolometric
corrections), since {\it LISA} estimates for the BH masses are
extremely precise by astronomical standards (see
Tab.~\ref{tab:LISA}). Second, if the quasar accretes at or near the
Eddington limit, given its Eddington ratio, one may be able to set a
useful lower limit on the radiative efficiency of its accretion flow.
If it is in excess of the canonical $10$\% value, it will provide an
interesting empirical test of the physics of accretion onto a spinning
BH, with a spin directly constrained by the GW measurement.  Third,
the counterpart could be monitored for years following the merger to
follow the viscous evolution of the gaseous disk and thus clarify its
role in the SMBH coalescence process. Fourth, it is expected that the
gas disk will be forced in the orbital plane of the pre-merger binary
by the Bardeen-Peterson effect \citep{mp05}. Knowing the disk
orientation could thus offer tests of the geometry of quasar emission
and obscuration (even after the merger, given the expected spin of the
remnant). It may also be possible to further develop diagnostics
related to the geometry of a jet, if present.

\subsection{Implications: Cosmology}

The successful identification of an EM counterpart to a GW event
could, in principle, open the way to use them as ``standard sirens''
to probe the background cosmology \citep{s86}, analogously to the Ia
SNe standard candles \citep{hh05}. The precision on the cosmological
model can, however, be improved only if the $\dL(z)$ function is determined to a higher
accuracy than it can be already guessed from other data that exists
when {\it LISA} is operational. As an example, we have assumed here
to have available the uncertainty from the combined datasets from
two future projects, {\it Planck} and LSST. We have found that the
major obstacle against a dramatic improvement on cosmology is the
gravitational lensing of intervening matter along the line of sight
to a {\it LISA} source.  In the weak-lensing limit, the r.m.s.
magnification of a source at $z=2$ is $12\%$, leading to a
luminosity distance error of $6\%$. \cite{dhcf03} have shown that
galaxy shear maps can be used to correct weak-lensing distortion,
but only a 20\% relative improvement can be achieved, so that
$\delta \dL/\dL =5\%$. We suggest alternatively that correcting for
the contribution of the known distribution of intervening matter
might improve weak lensing uncertainties by another 20\%.
Furthermore, the weak lensing uncertainty can be overcome if a large
sample of sources is available when fitting $\dL(z)$, mapping out
the magnification distribution. Using a number $K$ of merger events
along with uniquely identified counterparts, the lensing error
reduces approximately as $1/\sqrt{K}$ (although the actual
improvement will be less pronounced, due to a non-Gaussian tail of
high magnifications; \citealt{whm02,hl05}). To reach the {\it
Planck} + LSST level of $\delta\dL / \dL=10^{-3}$, we find that
$K>100$ events would be required.

Is such a large number of merger events, with uniquely determined
electromagnetic counterparts, expected from the {\it LISA}
data-stream? Monte-Carlo simulations of SMBH merger trees generally
indicate {\it LISA} event rates from $\sim 20$--$0.5{\rm yr}^{-1}$
for masses $M_{\rm BH}\lsim 10^7$ at $z\sim 1$
\citep{mhn01,ses04,eins05}. Detecting a total of $K=100$ would be
barely possible during a 3-year {\it LISA} mission lifetime, only
allowing a marginal test of the concordance of cosmological
parameters with standard sirens.  In addition, most of these events
may be expected to involve SMBHs at the low--mass end (i.e. $\lsim
10^5 {\rm M_\odot}$; \citealt{m03,ses04}), which are not ideal for
unique counterpart identifications (see Fig.~\ref{fig:cp1}).
However, large uncertainties remain on the expected event rate. For
example, \cite{its04} predict much larger rates, which could open up
the possibility of a statistical analysis, folding in the expected
weak lensing magnification distribution.

\section{Conclusions}
\label{sec:conclusions}

In this paper, we have considered the possibility that SMBH-SMBH
mergers, detected as gravitational wave sources by {\it LISA}, are
accompanied by gas accretion and quasar activity with a
luminosity approaching the Eddington limit.  Under this assumption, we
have computed the number of quasar counterparts that would be found in
the three--dimensional error volume provided by {\it LISA} for a given
GW event. We found that weak lensing errors exceed other sources of
uncertainties on the inferred redshift of the electromagnetic
counterpart and increase the effective error volume by nearly an order
of magnitude.  Nevertheless, we found that for mergers between
$\sim(4\times10^5 - 10^7) {\rm M_\odot}$ SMBHs at $z\sim 1$, the error
box may contain a single quasar with a B-band luminosity $L_B\sim
(10^{10}-10^{11}) {\rm L_\odot}$.  This would make the identification
of unique electromagnetic counterparts feasible, allowing precise
determinations of the Eddington ratio of distant accreting SMBHs, and
providing an alternative method to constrain cosmological parameters.

\acknowledgments

We thank Dan Holz, Scott Hughes, Szabolcs M\'arka, and Merse E.
G\'asp\'ar for useful discussions. We are grateful to Sheng Wang for
providing the Fisher matrices.  Z. H. was supported in part by NSF
through grants AST-0307200 and AST-0307291, by NASA through grant
NNG04GI88G, and by a Gy\"orgy B\'ek\'esy Fellowship from the
Hungarian Ministry of Education. Z. F. acknowledges support from
OTKA through grant nos. T037548, T047042, and T047244.

\end{document}